\documentstyle[11pt]{article}
\input FEYNMAN
\title{\vspace{-1.in} \hfill {\small\rm TUM-HEP-326/98} \\~\\~\\
New Physics with Mirror Particles}
\author{ George Triantaphyllou\thanks{e-mail:georg$@$ph.tum.de}
$\;$\\~  \\{\it Institut f\"ur Theoretische Physik, Technische 
Universit\"at M\"unchen}\\
{\it James-Franck-Strasse, D-85748 Garching, GERMANY }}
\begin{document}
\setlength{\baselineskip}{24pt}
\maketitle
\begin{abstract}
The introduction of mirror fermions  
with masses between the weak scale and 1 TeV could 
offer a dynamical origin to the standard-model
electro-weak symmetry  breaking mechanism.
The purpose of this work is to study the dynamics needed in order
to render models with such a
fermion content phenomenologically acceptable. 
\vspace{2.in}
\end{abstract}
\setcounter{page}{0}
\pagebreak

\section{Physical motivation}

The  standard-model description of 
the electro-weak interactions of elementary
particles is in general agreement 
with present experimental data. In this
model, the Higgs mechanism is implemented in order
to break spontaneously the $SU(2)_{L} \times
U(1)_{Y}$
gauge symmetry, to give fermion masses
and to unitarize the $W^{+}W^{-}$ scattering
amplitude. Naturalness arguments (the hierarchy problem)
have frequently led to studies
where this mechanism is just an effective low-energy
description of strong non-perturbative dynamics
involving new gauge and matter degrees of freedom,
in which case the Higgs particle is a composite state of new fermions
whose condensate breaks dynamically the electro-weak symmetry.

On the other hand, the proximity of the top-quark mass to the 
electro-weak scale suggests an active role of the top-quark to
such a condensate. However, relevant studies have shown that the 
top quark by itself is hardly heavy enough to reproduce the weak scale 
correctly, and in any case cannot eliminate the 
fine-tuning problem \cite{Manfred}.
A solution to this issue is to introduce new particles which mix with
the top quark and which are heavy enough to break the electro-weak
gauge symmetry
at the right scale. This not only eliminates the need for excessive fine
tuning which plagues the simpler top-condensate models, but also allows 
for a naturally heavy top quark. 

An attempt  along these lines was presented some time ago \cite{litri}
which involved mirror fermion generations, i.e. fermions
with interchanged isospin charges in a left-right 
symmetric context. However, that approach had several problems
that will be overcome here.
A first speculation for the existence of mirror fermions appeared
in the classical paper on parity violation \cite{Lee} that led to 
the V-A interaction models. Efforts to 
eliminate completely  mirror fermions from nature are for some reminiscent 
of efforts several decades ago to identify the anti-electron with
the proton, and amounts to not realising that particles consistent with  
natural symmetries could actually exist independently.
Such a gauge group and fermion extension,
apart from fitting nicely into unification schemata, 
restores in a certain sense
the left-right symmetry of the matter sector missing in the standard model
or in the simplest left-right symmetric models. 
In this way it also
provides a well-defined continuum limit of the theory, something 
which is usually problematic due to the Nielsen-Ninomiya theorem 
\cite{Nielsen}. 

The left-right symmetric approach to standard-model 
extensions renders the baryon-lepton number 
symmetry $U(1)_{B-L}$
more natural by gauging it, and has also been proposed
as a solution to the strong CP problem  when 
accompanied with the introduction of 
mirror fermions \cite{Barr}. As will be seen in the following, 
the model  introduced here proves further  
to be economical by identifying the source of the strong dynamics which 
break the electro-weak symmetry dynamically with
a ``horizontal" generation gauge group in the mirror sector
which, apart from preventing the pairing-up of the standard-model
generations with the mirror ones, provides also the
intra-generation mass hierarchies. 

Furthermore, it was recently shown \cite{georgnew} that the
gauge and fermion content of the present model
is consistent with superstring-inspired
unification schemata, including the mirror fermion generation group. 
The corresponding gauge coupling unification does not pose problems
with proton decay and allows
the prediction, in order of magnitude, of the QCD and weak scales. 
In such a superstring-inspired unification, possibly 
connected to $N=2$ supergravity, the standard-model fermions would 
have both mirror
and supersymmetric partners. The present approach corresponds to 
breaking  supersymmetry and leaving the
supersymmetric partners  close to the unification scale, and
bringing the mirror partners down to the weak scale, altering 
thus radically the expected phenomenology. 
 
From the experimental side,  mirror fermions at low
scales could already have manifested their existence indirectly.  
The nature of the non-perturbative interactions introduced could namely 
be related to the $3 \sigma$ deviation from 
the standard-model value of the right-handed bottom-quark weak coupling
extracted from the $A_{b}$ asymmetry \cite{Field}, as we will later see. 
If a similar anomaly is conjectured for the top quark, it could influence 
substantially the
values of the electroweak precision parameters $S$ and $T$ which, even though
experimentally still consistent with zero, 
have negative mean values and can be non-negligibly negative, signaling
new physics. 

On the other hand, 
the introduction of mirror fermions could  also  create  problems with
the $S$ parameter, which  can receive large positive 
``oblique" corrections due to
12 new chiral fermion doublets, as we will see in section 3. 
This parameter could of course also receive  negative contributions from 
the vertex corrections directly related to  the anomalous
couplings mentioned above. Nevertheless, the same vertex effects that are 
taken to cancel the ``oblique" corrections could potentially 
have a different sign, adding to these effects instead and rendering the
model phenomenologically unacceptable. 
Their non-perturbative nature does not allow unfortunately 
the {\it a priori} determination of the sign and magnitude of their
contributions, and the working assumption in this work will be that they take 
values in agreement with present bounds on the electroweak precision
parameters.

In the following section, the dynamics needed to make such a
mechanism phenomenologically viable are carefully analyzed. 
In particular, it  proves  necessary to review 
the dynamical assumptions made in Ref.\cite{litri}.  
In that work it was unclear why the characteristic scale
of the strong group responsible for the fermion gauge-invariant masses
happened to be so close to the scale where 
the strong interactions breaking electro-weak symmetry 
became critical. Furthermore, the
previous model 
could not provide a see-saw mechanism for the standard-model neutrinos,
coupling unification would be difficult,  
it had problems with the isospin quantum numbers of the lighter fermions,  
and it needed fine-tuning in order to prevent some fermions from 
acquiring large masses.

 In the present approach, 
 only the mirror particles are coupled strongly and 
 dynamically involved in the 
breaking of $SU(2)_{L}$. By eventually breaking the mirror-generation
symmetries,  small gauge-invariant (by this we mean here and in the 
following gauge-invariant 
under the standard-model gauge group, unless otherwise stated)
masses are allowed 
which communicate the electro-weak symmetry breaking to the 
standard-model fermions by mixing them with their mirror partners. 
This model has neither ``sterile" nor $SU(2)_{L}$-doublet  
light mirror neutrinos, as in \cite{Bag} for example, 
which would pose problems with experiment. 
After the mass hierarchies are computed within this context, phenomenological
consequences like electro-weak precision parameters, CKM
 matrix elements, flavor-changing neutral currents (FCNC) and decays
 are discussed.
  
\section{The model}
The gauge-group structure considered is described by
$SU(4)_{PS} \times SU(2)_{L} \times SU(3)_{2G} 
\times SU(2)_{R}$.
The group $SU(4)_{PS}$ is the usual Pati-Salam group unifying
quarks and leptons, and $SU(2)_{L}$ is the group of weak interactions.
The group $SU(3)_{2G}$ is a horizontal gauge symmetry 
acting only on the mirror
fermions, which becomes strong at around 2 TeV. All other groups are
taken to have weak couplings at this energy. 
The corresponding symmetry for the standard-model fermions  
$SU(3)_{1G}$ 
 has already been broken at higher scales, at once or 
 sequentially, in order to avoid large FCNC.  

  Under the above gauge structure, 
  the following fermions are introduced, which are left-handed gauge (and not
  mass) eigenstates and transform like 
\begin{displaymath}
\begin{array}{clcl} 
&{\rm Generations} && {\rm Mirror \;\;\; generations}\\~ \\
\psi_{L}:& ( {\bf 4, \;2, \;1, \;1}) 
& \psi^{M}_{L}:&( {\bf 4, \;1, \;3, \;2}) 
\\
\psi_{R}^{c}:& ( {\bf \bar{4}, \;1,  \;1, \;2} ) 
& \psi^{M\;c}_{R}:& ( {\bf \bar{4}, \;2,  \;3, \;1} )   
\end{array}
\nonumber
\end{displaymath}

  \noindent 
The superscript $M$ denotes the mirror partners
of the ordinary fermions, and $c$ denotes charge conjugation.

  One observes that the generation symmetries play a very
  important role at this stage, and this is to  prevent the formation of 
large  gauge-invariant masses. Pairing-up of standard-model and mirror
 generations is thus prohibited, in agreement with what is usually
 called ``survival hypothesis" \cite{Howard}.

 Even though this 
 quantum number assignment is reminiscent of technicolor with
 a strong group 
 $SU(N)_{TC} \approx SU(3)_{2G}$, there is no corresponding
 extended technicolor (ETC) group,
there is a left-right interchange of weak isospin  charges, and 
the new anti-particles transform under the same 
(and not the complex conjugate) representation of the
strong group as the new particles.  
Even though the latter difference is not essential for the 
present work, it is introduced for two reasons. The first one
is that leaving
a possibility to the strong mirror group to self-break even partially
could be useful to the subsequent theoretical development of the model. 
The second is that it can fit easier in unification groups emerging in  
some superstring models \cite{georgnew}.  Possible remaining anomalies 
related to this matter content are assumed to be cancelled by new physics,
like additional fermions, at unification scales.
 The breaking of the strong
 group in the present case is an additional difference from
 technicolor models.
 One should furthermore not  confuse the 
present model with other ``mirror" fermion approaches, like in
\cite{Mex} for example, where all
components of the new fermions are singlets under $SU(2)_{L}$
and interact only gravitationally or marginally with the
standard-model particles, and
which obviously cannot break the electro-weak symmetry dynamically. 

\subsection{Getting to the standard model}

At  high energy  scales that do not enter directly
in this work, the Pati-Salam group is assumed to break spontaneously like 
$SU(4)_{PS} \times SU(2)_{R}  
\longrightarrow SU(3)_{C}\times  U(1)_{Y}$, where $SU(3)_{C}$ and
$U(1)_{Y}$ are the usual QCD and hypercharge groups respectively. 
(The particular way of the Pati-Salam-group breaking does not influence the
discussion that follows. For alternative ways, see Ref.\cite{georgnew}.)  
Much later, at scales on the order of $\Lambda_{G} \approx 2$ TeV,   
the mirror generation group breaks sequentially, 
just after it becomes strong, like   
$SU(3)_{2G} \longrightarrow
SU(2)_{2G}
\longrightarrow \emptyset$, 
It is not attempted here to investigate
how exactly these breakings occur, and for simplicity it is 
enough to assume that
a Higgs mechanism is responsible for them, effective or not.   

The first spontaneous
generation symmetry breaking $SU(3)_{2G}
\longrightarrow SU(2)_{2G}$  occurs at
a scale $\Lambda_{G}$, with an 
$SU(2)_{L}$-singlet scalar state denoted by $\phi_{3}$ 
and transforming like a ${\bf 3}$ under 
the generation symmetry acquiring a non-zero vev. 
Note that the group $SU(3)_{2G}$ could in 
principle partially self-break dynamically  
{\it via} the fermion-condensation
channel ${\bf 3} \times {\bf 3} \longrightarrow 
{\bf \bar{3}}$ if it were given the chance 
to become strongly coupled at this energy scale. 
We comment later on the possible connection of this channel
to the large top-quark mass.

The fermions have the 
following quantum numbers under the new gauge symmetry 
$SU(3)_{C} \times SU(2)_{L} \times SU(2)_{2G}
 \times U(1)_{Y}$:   

\begin{displaymath}
\begin{array}{clcl} 
&{\rm The \;\;3rd\;\;\&\;\; 2nd \;\;generations} 
&& {\rm The \;\;3rd\;\& \;\;2nd \;\;mirror 
\;\; generations} \\~ \\
q^{3,2}_{L}:& ( {\bf 3, \;2,  \;1,} \;1/3  ) 
& q^{3,2M}_{L}:&( {\bf 3, \;1,  \;2,}  \;~^{+4/3}_{-2/3} ) 
\\~ \\ 
l^{3,2}_{L}:& ( {\bf 1, \;2,  \;1,}  \;-1  ) 
& l^{3,2M}_{L}:& ( {\bf 1, \;1,  \;2,} \;~^{\;0}_{-2} ) 
\\~ \\ 
q^{3,2\;\;c}_{R}:& ( {\bf \bar{3}, \;1,  \;1,} 
\;~^{-4/3}_{+2/3} ) 
& q^{3,2M\;\;c}_{R}:& ( {\bf \bar{3}, \;2,  \;2,} 
 \;-1/3 )   
\\~ \\
l^{3,2\;\;c}_{R}:& ( {\bf 1, \;1,  \;1,} \;~^{0}_{2} ) 
& l^{3,2M\;\;c}_{R}:& ( {\bf 1, \;2,  \;2,}
 \;1 )  \\~\\~\\ 
\end{array}
\end{displaymath}
\begin{displaymath}
\begin{array}{clcl} 
&{\rm The \;\;1st \;\;generation\;\;\;\;\;\;\;\;\;\;\;\;} 
&& {\rm The \;\;1st \;\;mirror 
\;\; generation\;\;\;\;\;\;\;\;\;} \\~ \\
q^{1}_{L}:& ( {\bf 3, \;2,  \;1,}  \;1/3  ) 
& q^{1M}_{L}:&( {\bf 3, \;1,  \;1,}  \;~^{+4/3}_{-2/3} ) 
\\~ \\ 
l^{1}_{L}:& ( {\bf 1, \;2,  \;1,}  \;-1  ) 
& l^{1M}_{L}:& ( {\bf 1, \;1,  \;1,} \;~^{\;0}_{-2} ) 
\\~ \\ 
q^{1\;\;c}_{R}:& ( {\bf \bar{3}, \;1,  \;1,} 
\;~^{-4/3}_{+2/3} ) 
& q^{1M\;\;c}_{R}:& ( {\bf \bar{3}, \;2,  \;1,}  \;-1/3 )   
\\~ \\
l^{1\;\;c}_{R}:& ( {\bf 1, \;1,  \;1,}  \;~^{0}_{2} ) 
& l^{1M\;\;c}_{R}:& ( {\bf 1, \;2,  \;1,}  \;1 )  \\~\\~\\ 
\end{array}
\end{displaymath}

\noindent  where the superscripts 1,...,3 indicate the fermion
generations.
 Moreover, the letters  $q$ and $l$ stand for 
quarks and leptons respectively.
Note that $\bar{\psi_{R}}\psi^{M}_{L}$ mass terms are still
prohibited by the $SU(2)_{2G}$ symmetry for the second and third
generations.

At a scale quite close to 
$\Lambda_{G}$,  the $SU(2)_{2G}$ 
group breaks spontaneously sequentially to $U(1)_{G}$ and this
down to $\emptyset$ by  
two $SU(2)_{L}$-singlet 
scalar states, denoted by $\phi_{2,1}$ and  
transforming like a ${\bf 2}$ and  being charged 
respectively under the generation
symmetries, which  acquire  non-zero vevs.
The quantum numbers of the third and second generation mirror fermions 
after these breakings are given by
\begin{displaymath}
\begin{array}{clcl} 
&{\rm The \;\;2nd\;\;mirror\;\;generation} && {\rm The \;\;3rd \;\;mirror 
\;\; generation} \\~ \\
 q^{2M}_{L}:&( {\bf 3, \;1,} \;~^{+4/3}_{-2/3} ) 
& q^{3M}_{L}:&( {\bf 3, \;1,}  \;~^{+4/3}_{-2/3} ) 
\\~ \\ 
 l^{2M}_{L}:& ( {\bf 1, \;1,} \;~^{\;0}_{-2} ) 
& l^{3M}_{L}:& ( {\bf 1, \;1,} \;~^{\;0}_{-2} ) 
\\~ \\ 
 q^{2M\;\;c}_{R}:& ( {\bf \bar{3}, \;2,} 
 \;-1/3 )   
& q^{3M\;\;c}_{R}:& ( {\bf \bar{3}, \;2,} \;-1/3 )   
\\~ \\
 l^{2M\;\;c}_{R}:& ( {\bf 1, \;2,} \;1 )  
& l^{3M\;\;c}_{R}:& ( {\bf 1, \;2,} \;1 )  \\~\\~\\ 
\end{array}
\end{displaymath}

\noindent while the first mirror generation and all the standard-model 
generation quantum numbers are left unchanged. 
 
The breakings of the mirror generation symmetries   
described above induce at lower energies, among others, effective
four-fermion operators $F$ of the form 
\begin{eqnarray}
F &= & \frac{\lambda}{\Lambda^{2}_{G}}
({\bar \psi^{M}_{R}}\psi^{M}_{L})({\bar \psi^{M}_{L}}\psi^{M}_{R})  
\end{eqnarray}

 \noindent 
 for the three mirror fermion generations, where $\lambda$ 
 are effective four-fermion couplings and the
generation indices are omitted for simplicity. 
The fermion bilinears in parentheses above
 transform like  doublets under $SU(2)_{L}$. 

The next step is to assume that, 
in a manner analogous to top-color scenarios \cite{Hill},    
the $SU(2)_{2G}$ group
is strongly coupled just before it 
breaks,  and it is therefore plausible to take    
the effective four-fermion couplings $\lambda$
 to be critical for the corresponding mirror generations, 
 like in the Nambu-Jona-Lasinio model (NJL).    
Therefore, condensates of mirror fermions like 
$<{\bar \psi^{M}_{L}}\psi^{M}_{R}>$   
 can form which  break the symmetry 
$SU(2)_{L}\times  U(1)_{Y}$ dynamically down to
the usual $U(1)_{EM}$ group of electromagnetism. 

\subsection{The mass generation}

The 
fermion condensates described above  give to the mirror fermions 
symmetry-breaking
masses of order $M  \approx r \Lambda_{F}$  
 {\it via} the operators $F$,   
with $r$  a constant not much smaller than unity if
one wants to avoid excessive fine-tuning of the four-fermion interactions. 
Effective operators of the form 
${\bar \psi^{1\;M}_{R}}\psi^{1\;M}_{L}
{\bar \psi^{2,3\;M}_{L}}\psi^{2,3\;M}_{R}/\Lambda^{2}_{G}$ 
induced by the broken $SU(3)_{2G}$ interaction  feed
down gauge-symmetry-breaking masses to the first 
mirror generation. The fact that all mirror fermions get large masses
of the same order of magnitude 
due to the critical interactions avoids fine-tuning problems that
would appear if mass hierarchies were introduced by allowing only some of 
them to become massive, as is done in \cite{Georgi}. 
Moreover, to avoid breaking QCD and 
electromagnetism, it is assumed that most-attractive-channel
dynamics prevent quark-lepton 
condensates of the form $<\bar{q^{M}_{L}}l^{M}_{R}>$
from  appearing. 

If generation symmetries were left intact, the mass matrix ${\cal M}$
for all the fermions would have the form  
\begin{eqnarray}
&& \;\;\; \psi_{L} \;\;\;\; \psi^{M}_{L} \nonumber \\ 
\begin{array}{c} {\bar \psi}_{R} \\ {\bar \psi^{M}}_{R} \end{array} && 
\left(\begin{array}{cc} 0 & \;\;\;0 \\ 0 & 
\;\;\;M \end{array} \right), 
\nonumber
\end{eqnarray} 
\noindent where the 4 elements shown are blocks of $3\times3$ 
matrices in generation space and $M$ the dynamical mirror-fermion
mass due to the strong generation interactions. 
However, the broken 
generation symmetries allow the formation of gauge-invariant 
masses, and the mass matrix ${\cal M}$ takes the form:
\begin{eqnarray}
&& \;\;\; \psi_{L} \;\;\;\;\;\;\; \psi^{M}_{L} \nonumber \\ 
\begin{array}{c} {\bar \psi}_{R} \\ {\bar \psi^{M}}_{R} \end{array} && 
\left(\begin{array}{cc} 0 & \;\;\;m_{1} \\ m_{2} & 
\;\;\;M \end{array} \right), 
\nonumber
\end{eqnarray}

\noindent  where the diagonal elements  
are gauge-symmetry breaking and the off-diagonal 
gauge-invariant. 

The off-diagonal mass matrices can be generated by
Yukawa couplings $\lambda_{ij}$ associated with spinor
bilinears of fermions with their mirror partners which are
coupled to  the scalar states $\phi_{2,3}$ 
responsible for the  
spontaneous generation symmetry breakings. The corresponding  
gauge-invariant term in the Lagrangian  has the form 
$\sum_{i,j} \lambda_{ij} \bar{\psi_{iR}}\psi^{M}_{jL} \phi_{2,3}$, where
the indices $i,j$ count the corresponding fermions in the model. The
elements of the matrices $m_{1,2}$ will be taken in general to be
quite smaller than
the ones in the matrix $M$, with the exception of the entries
related to the top quark. 

After diagonalization of the mass matrix shown above, in which the lighter
mass eigenstates are identified with the standard-model fermions,
a  {\em see-saw} 
mechanism  produces  small masses for the
ordinary fermions and  larger ones 
for  their mirror partners. A specific 
example for illustration purposes is produced in the next section. 
The situation is reminiscent of universal see-saw models, but 
it involves fermions having quantum-number assignments which should not in
principle pose problems with the Weinberg angle 
$\sin^{2}{\theta}_{W}$ \cite{Bur}, \cite{Cho}.  

Some  remarks relative to the (1,1) block entry of the mass matrix
are in order. First, there are no $<{\bar \psi_{R}}\psi_{L}>$
condensates at these high energy scales. 
Then, after careful inspection
of the quantum numbers carried by the gauge bosons of the broken groups
one observes that there are no four-fermion effective operators
of the form 
$({\bar \psi_{R}}\psi_{L})({\bar \psi^{M}_{L}}\psi^{M}_{R})/\Lambda^{2}_{G}$ 
or any other gauge-invariant operators
for any generation which would feed
gauge-symmetry-breaking  
masses to the ordinary fermions at this stage.

\section{Phenomenology} 

\subsection{Masses and mixings} 

We start by calculating the mass hierarchies produced by the model, since
they provide the basis of any phenomenological analysis. 
The gauge-symmetry breaking mass submatrices $M$ 
are hermitian because of parity symmetry.
 The gauge-invariant ones, denoted by $m_{1,2}$ should be symmetric
 due to the quantum numbers assigned to the fermions,   
 but not necessarily real. 
Complex matrix elements allow therefore
in general  for weak CP violation.
Assuming that $SU(2)_{L}$ effects can be  
neglected in the gauge-invariant mass generation process or
that their effect is just homogeneously multiplicative, one also has 
the relation $m_{2}= c\,m^{\dagger}_{1}$ between the
gauge-invariant submatrices, with $c$ a real constant. 
This  means that the determinant of the mass matrix ${\cal M}$ is real, 
eliminating thus the strong CP problem in this approximation,
at least at tree level.

For simplicity, the mass matrices in the following are  taken real and
having the form 
\begin{eqnarray}
&& {\cal M}_{i} = \left(\begin{array}{cc}  0 & m_{i} \\ m_{i} & M_{i} 
 \end{array} \right), i = U,D,l    
\end{eqnarray}

\noindent for the up-type quarks ($U$), down-type quarks ($D$) and charged
leptons ($l$). We give as a numerical example 
forms for the off-diagonal gauge-invariant 
mass submatrices of the up-type and down-type quark
sectors for illustration purposes (with obvious correspondence
between column and row numbers with generation indices):   
\begin{eqnarray}
\begin{array}{c}  \\ m_{U} ({\rm GeV}) =\\ \end{array} &&
\left(\begin{array}{ccc} 2.3 & 5.7 & 1.1 \\ 5.7 & 20 & 1.3 \\
1.1 & 1.3 & 360 \end{array} \right)  
\begin{array}{c}  \\,\;\;  m_{D} ({\rm GeV}) =\\ \end{array} 
\left(\begin{array}{ccc} 1.6 & 1.6 & 0.51 \\ 1.6 & 4 & 1.3 \\
0.51 & 1.3 & 35 \end{array} \right).    
\end{eqnarray}

\noindent
The dynamical assumption is made here that the $SU(2)_{L}$-breaking 
mass submatrices are diagonal and have the form 
\begin{eqnarray}
\begin{array}{c}  \\ M_{U} ({\rm GeV}) =\\ \end{array} &&
\left(\begin{array}{ccc} 360 & 0 & 0 \\ 0 & 650 & 0 \\
0 & 0 & 650 \end{array} \right)  
\begin{array}{c}  \\,\;\;  M_{D} ({\rm GeV}) =\\ \end{array} 
\left(\begin{array}{ccc} 200 & 0 & 0 \\ 0 & 360 & 0 \\
0 & 0 & 360 \end{array} \right).  
\end{eqnarray}

\noindent The gauge-symmetry breaking masses of the first mirror generation
are taken to be smaller than the ones of the two heavier
generations corresponding to $SU(2)_{2G}$. It should be noted here 
that, in principle, $SU(2)_{2G}$ could equally well correspond to the
mirror partners of the two lighter fermion generations. 
Such a scenario presents a particular interest since the heaviness of
the top quark could be directly related to the self-breaking channel 
${\bf 3} \times {\bf 3} \longrightarrow {\bf \bar{3}}$ 
of $SU(3)_{2G}$ mentionned before (the completion of the 
generation-group breaking being attributed to QCD for instance,
eliminating thus the need for elementary scalars).

 It is  also expected that the dynamics provide some 
 custodial symmetry breaking which is responsible for 
 the mass difference in the up- and down-quark sectors. The $U(1)_{Y}$ 
 could be in principle the source of this difference, but 
 we do not speculate on how this is precisely realised here. 
One has to further stress that the splitting of $M_{U}$ and
$M_{D}$ is not {\it a priori} needed to produce the
top-bottom quark mass hierarchy , but it is introduced only to
better fit the experimental constrains on the 
electro-weak parameters, as  will be seen later.

These mass matrices give, after diagonalization and without the need for
any fine-tuning,  the following
quark and mirror-quark masses  (given in units of GeV):  
 
\noindent 
Standard-model quarks \hfill Mirror quarks $\;$ \\ 
$m_{t} = 160$, $m_{c} = 0.77$, $m_{u} = 0.001$  \hfill 
$m_{t^{M}} = 810$, $m_{c^{M}} = 651$, $m_{u^{M}} = 360\;$  \\
$m_{b} = 3.4\;$, $m_{s} = 0.07$, $m_{d} = 0.003$  \hfill 
$m_{b^{M}} = 363$, $m_{s^{M}} = 360$, $m_{d^{M}} = 200$. 
\newline
The ordinary quark masses given are slightly smaller than
the ones usually quoted because the values reported here are relevant
to the characteristic scale of the new strong dynamics which is around 
2 TeV, and one has therefore to account for their running with
energy. The formalism 
presents no inherent difficulty whatsoever producing larger masses
for these fermions. 

The generalization of the standard-model CKM quark-mixing matrix in this 
scenario is a unitary 
$6\times 6$ matrix of the form $V_{G}=K^{T}_{U}K_{D}$, with $K_{U,D}$
the linear operators diagonalizing the mass matrices of the up and
down fermion sector.  The generalized 
CKM matrix has the form 
\begin{eqnarray}
 V_{G}= \left(\begin{array}{cc} V_{CKM} & \;\;\; V_{1} \\ V_{2} & 
\;\;\; V^{M}_{CKM} \end{array} \right),  
\end{eqnarray} 

\noindent and the usual standard-model CKM matrix $V_{CKM}$ is one of its 
submatrices given (in absolute values) by 
\begin{eqnarray}
\begin{array}{c}  \\ | V_{CKM} | =\\ \end{array} &&
\left(\begin{array}{ccc} 0.98 & 0.22 & 0.003 \\ 0.22 & 0.97 & 0.042 \\
0.006 & 0.038 & 0.95 \end{array} \right),   
\end{eqnarray}

\noindent which is consistent with present experimental
constraints. 

The mixing between the first and second generations is larger
than the one between the second and third generations, and this
can be easily traced back to the relative elements of $m_{U,D}$. 
Furthermore, one has to be particularly cautious 
when using the flavor symbol `t' and the flavor name `top quark'
for the heaviest standard-model-quark mass eigenstate, since  
$t_{L} (t^{c}_{R})$ has a non-negligible $SU(2)_{L}$ 
singlet (doublet) component as expected due to the large
$\bar{t_{R}}t^{M}_{L}=\bar{t^{M}_{R}}t_{L}$ mass terms, and this 
is reflected on the reported value
of $|V_{tb}|=0.95$. This is particularly apparent in the
third-generation fermions to which correspond larger gauge-invariant masses, 
since the fermion-mirror fermion 
mixings are given  roughly by the ratio $m_{ii}/M_{ii}$.  
Present experimental data give $|V_{tb}|=0.99 \pm 0.15$ \cite{Tarta}. 
More precise future measurements of
this quantity should show deviations from its standard-model value which
is very close to 1 assuming unitarity of the mixing matrix $V_{CKM}$.  
Larger mirror-fermion masses can diminish this effect by
reducing the corresponding mixing of the mirrors with the
ordinary fermions. 

In fact, indirect experimental indications for the
existence of $SU(2)_{L}$-singlet
new fermions which can mix with the third standard-model-generation 
charged fermions
$t_{L}$, $b_{L}$, $\tau_{L}$, and $SU(2)_{L}$-doublet 
new anti-fermions which can mix 
with $t^{c}_{R}$, $b^{c}_{R}$ and $\tau^{c}_{R}$ 
could already exist in LEP/SLC precision data. 
One would be coming from the $S$ and $T$ parameters, which are 
consistent with anomalous right-handed
top-quark couplings, as will be seen later, 
and the other coming from anomalous right-handed $b$-quark and $\tau$-lepton
couplings  to the $Z^{0}$ boson corresponding to even $3 \sigma$ effects 
\cite{Field}. These are extracted from the current $A_{b,\tau}$ 
asymmetries. Note that, contrary to extended technicolor models that
introduce left-handed anomalous couplings and affect mostly $R_{b}$, 
this mirror model induces anomalous right-handed couplings which
affect mostly $A_{b}$.
The actual sign of the deviations depends on the relevant interaction 
strength of the two isospin partners of the mirror doublets with
the standard-model fermions, but more details
on  this are given in subsection 3.2. Deviations from  the weak couplings 
of the lighter standard-model particles are largely suppressed,  
but they can be potentially large when the mirror partners are light. 
Bringing all the mirror partners to lower scales 
should be avoided nevertheless, since reproducing the weak scale
would then require fine-tuning, as will be shown shortly.

The corresponding CKM matrix for the 
mirror sector $V^{M}_{CKM}$ is equal (in absolute values) to 
\begin{eqnarray}
\begin{array}{c}  \\ | V^{M}_{CKM} | =\\ \end{array} &&
\left(\begin{array}{ccc} 1 & 0.001 & 0.001 \\ 0.001 & 1 & 0.039 \\
0.001 & 0.036 & 0.95 \end{array} \right).    
\end{eqnarray}

\noindent 
The third generation is here the main reason  why this matrix is not diagonal
(The entries (1,1) and (2,2) are close to unity because
of the assumed diagonal form of $M_{U,D}$, but not exactly unity, so  
that the unitarity character of the mixing matrix $V_{G}$ is preserved.) 
Furthermore, the matrices 
$V_{1}$ and $V_{2}$ mix the up-quark sector of the 
standard model with the down mirror-quark 
sector and {\it vice-versa}, and are given (in absolute values) by 
\begin{eqnarray}
\begin{array}{c}  \\ | V_{1} |=\\ \end{array} &&
\left(\begin{array}{ccc} 
0.0037 & 0.0015 & 0.0002 \\ 
0.0070 & 0.0202 & 0.0027 \\
0.0005 & 0.0153 & 0.3163  
\end{array} \right)  
\begin{array}{c}  \\, | V_{2} |=\\ \end{array} 
\left(\begin{array}{ccc} 
0.0052 & 0.0060 & 0.0007 \\ 
0.0059 & 0.0193 & 0.0024 \\
0.0026 & 0.0163 & 0.3162 
\end{array} \right).  
\end{eqnarray}

The elements of these matrices 
are very small, apart from $|V_{tb^{M}}| = |V_{t^{M}b}|$
which account for the smallness of $|V_{tb}|$. This is  expected,
since, apart from the top quark,
the gauge-invariant masses responsible for the mixing
are much smaller than the gauge-symmetry breaking masses. 
One cannot expect therefore to find observable FCNC effects involving
the first two generations, like deviations in the $K_{L}-K_{S}$
meson mass difference. 
Processes involving the third generation however 
like $b \longrightarrow s \gamma$
could be affected and relevant deviations could be detectable in the future. 

One can investigate the predictive power of such a framework by 
comparing the number of input and output parameters required to produce
the numbers presented above. 
With 16 different input parameters 
one gets 12 masses (6 for the ordinary and 6 for
the mirror fermions) and 36 different mixing angles of the generalized CKM
matrix, which offers an advantage to the above considerations. Deeper 
insights into the gauge-invariant mass-generation mechanism should in 
the future further reduce the initial independent parameters.

For the charged leptons,  a diagonal gauge-symmetry 
breaking mass matrix  is used again and  a
gauge-invariant mass matrix having the  forms    
\begin{eqnarray}
\begin{array}{c}  \\ M_{l} ({\rm GeV}) =\\ \end{array} &&
\left(\begin{array}{ccc} 180 & 0 & 0 \\ 0 & 200 & 0 \\
0 & 0 & 200 \end{array} \right)  
\begin{array}{c}  \\ , \; m_{l} ({\rm GeV}) =\\ \end{array}
\left(\begin{array}{ccc} 0.25 & 0.25 & 0.1 \\ 0.25 & 3.8 & 1 \\
0.1 & 1 & 17 \end{array} \right).  
\end{eqnarray}

\noindent These give the following lepton and mirror-lepton 
mass hierarchy (at 2 TeV and in GeV units): \\  
Standard-model charged leptons \hfill Mirror charged leptons $\;$ \\ 
$m_{\tau} =1.45$ , $m_{\mu}=0.07$,   
$m_{e}=3\times10^{-4}$  \hfill   
$m_{\tau^{M}} = 201$,  
$m_{\mu^{M}} = 200$, $m_{e^{M}} = 180$. \newline 
The difference of the charged-lepton mass matrix with the
down-quark mass matrix is attributed to QCD effects.
The same mass hierarchies could have been produced with
a diagonal submatrix $m_{l}$ which would require less parameters, 
but for the sake of consistency a submatrix form similar to $m_{D}$
is chosen.
The neutrino mass and mixing
matrix is quite interesting and
is studied elsewhere \cite{georgnew}, since the fact that neutrinos 
can have both Dirac and Majorana masses makes theoretical 
considerations and calculations more involved.

At this point it is not claimed that  
the mass-matrix elements given above can be calculated explicitly 
 within this model, since these could in principle
receive important non-perturbative contributions. We just want to 
illustrate that it is feasible in principle
within this context to generate the correct
mass hierarchies and CKM angles.  
 Having now these ingredients allows us to tackle
 various other phenomenological issues.

\subsection{The weak scale and the electro-weak precision data} 

We next proceed by giving an estimate
for the dynamically generated weak scale $v$. 
A rough calculation using the Pagels-Stokar formula gives 
\begin{equation}
v^{2} \approx \frac{1}{4\pi^{2}} 
\sum^{N}_{i} M_{i}^{2} \ln{(\Lambda_{G}/M_{i})} \;, 
\end{equation}

\noindent where $N$ is the number of new weak doublets introduced and
$M_{i}$ their mass, where it has been  assumed for simplicity that
$m_{\nu^{M}_{i}} = m_{u^{M}}$ for all mirror neutrinos and where
departures from pure weak eigenstates have been neglected. 
Consequently, 
for the masses found before and  $\Lambda_{G} \approx $ 1.8 TeV
 one gets $v \approx 250$ GeV, as is required.  
The mirror fermions can therefore be heavy enough to eliminate any
need for excessive fine-tuning of the four-fermion interactions responsible
for their masses. This numerical example should not be taken at 
face value of course, since moderately heavier mirror fermions are still 
possible and  render even smaller values for $\Lambda_{G}$ acceptable. 

The $S$ parameter \cite{Tatsu} could be problematic in this scenario
however, since 
12 new $SU(2)_{L}$ doublets are introduced. 
 The main negative effect able to cancel the corresponding
 large positive contributions to $S$ coming from ``oblique" corrections 
 is the existence of vertex corrections 
 stemming from 4-fermion
effective interactions, which can give rise to  similar effects as   
the ones induced by  light $SU(2)_{L}$-invariant scalars 
known as ``techniscalars" \cite{Kagan}.    

More precisely, it is argued that the effective Lagrangian of the theory 
contains terms which can lead to a shift to the couplings of the top and 
bottom quarks to the $W^{\pm}$ and $Z^{0}$ bosons.   
In particular, there are four-fermion terms
involving 3rd generation-quark flavor eigenstates and their
mirror partners given by  
\begin{eqnarray}
{\cal L}_{{\rm eff}} =& -& \left(
\frac{\lambda_{n1}}{\Lambda^{2}_{n1}} 
\bar{t^{M}_{L}}\gamma^{\mu}t^{M}_{L}
+ \frac{\lambda_{c1}}{\Lambda^{2}_{c1}}
\bar{b^{M}_{L}}\gamma^{\mu}b^{M}_{L}
\right)
\bar{t_{R}}\gamma_{\mu}t_{R} \nonumber - \\   
&-&\left( 
\frac{\lambda_{c2}}{\Lambda^{2}_{c2}} 
\bar{t^{M}_{L}}\gamma^{\mu}t^{M}_{L}
+ \frac{\lambda_{n2}}{\Lambda^{2}_{n2}} 
\bar{b^{M}_{L}}\gamma^{\mu}b^{M}_{L}
\right) \bar{b_{R}}\gamma_{\mu}b_{R}
\end{eqnarray}

\noindent where the $\lambda$'s and
$\Lambda$'s are
the effective positive couplings and scales of the 
corresponding operators renormalized at the $Z^{0}$ boson mass, and
the subscripts $n,c$ indicate whether the participating fermions have
the same hypercharge or not. One should at this point furthermore note that
terms like   $\frac{\lambda_{n3}}{\Lambda^{2}_{n3}}  
(\bar{q^{M}_{R}}\tau^{a}\gamma^{\mu}q^{M}_{R}) 
(\bar{q_{L}}\tau^{a}\gamma_{\mu}q_{L})$,  
where  $\tau^{a}, a=1,2,3$ are the three $SU(2)_{L}$ generators, 
cannot be generated here in perturbation theory,
unlike analogous terms in extended technicolor models. Anyway,
such terms would 
produce shifts only to the left-handed fermion couplings, and these 
are already too much constrained from LEP/SLC 
data to be of any interest here. 

Adopting the effective Lagrangian approach for the 
heavy, strongly interacting sector of the theory \cite{Sekhar},  
the two mirror-fermion currents are expressed in terms of effective 
chiral fields $\Sigma$ like  
\begin{eqnarray}
 \bar{t^{M}_{L}}\gamma^{\mu}t^{M}_{L}  
&=& i \frac{v^{2}}{2}{\rm Tr}\left(\Sigma^{\dagger}
\frac{1+\tau^{3}}{2}D^{\mu}\Sigma \right) 
 \nonumber \\ 
 \bar{b^{M}_{L}}\gamma^{\mu}b^{M}_{L} 
&=& i \frac{v^{2}}{2}{\rm Tr}\left(\Sigma^{\dagger}
\frac{1-\tau^{3}}{2}D^{\mu}\Sigma \right) 
\end{eqnarray}

\noindent where the covariant derivative $D^{\mu}$ is defined by  
\begin{equation}
D^{\mu}\Sigma = \partial^{\mu}\Sigma +ig \frac{\tau^{a}}{2}W_{a}^{\mu}\Sigma 
-ig^{\prime} \Sigma \frac{\tau^{3}}{2}B^{\mu}.   
\end{equation}

\noindent 
The $g$ and $g^{\prime}$ above are the couplings
corresponding to the gauge fields $W^{\mu}_{a}$ and $B^{\mu}$ of
the groups $SU(2)_{L}$ and $U(1)_{Y}$ respectively.
The chiral field 
$\Sigma=e^{2i \tilde{\pi}/v}$ transforms like $L\Sigma R^{\dagger}$
with $L \in SU(2)_{L}$ and $R \in  U(1)_{Y}$ as usual, 
with hypercharge $Y=\tau^{3}/2$ and
$\tilde{\pi}= \tau^{a}\pi^{a}/2$ containing the would-be
Nambu-Goldstone modes $\pi^{a}$ ``eaten" by the electro-weak bosons. 

In the unitary gauge one takes $\Sigma=1$, and the currents given 
above induce shifts in the standard-model Lagrangian
of the form
\begin{eqnarray}
\delta {\cal L}&=&(gW^{\mu}_{3}-g^{\prime}
B^{\mu})\left ( \delta g^{t}_{R}
\bar{t_{R}}\gamma_{\mu}t_{R} +
\delta g^{b}_{R}\bar{b_{R}}\gamma_{\mu}b_{R} \right) 
\end{eqnarray} 

\noindent 
with the non-standard fermion-gauge boson couplings expressed by  
\begin{eqnarray}
\delta g^{t}_{R} &=& \;\; \frac{v^{2}}{4}
\left(\frac{\lambda_{n1}}{\Lambda^{2}_{n1}}-
\frac{\lambda_{c1}}{\Lambda^{2}_{c1}}\right) \nonumber \\ 
\delta g^{b}_{R} &=& 
- \frac{v^{2}}{4}
\left(
\frac{\lambda_{n2}}{\Lambda^{2}_{n2}} 
-\frac{\lambda_{c2}}{\Lambda^{2}_{c2}} \right). 
\end{eqnarray}

\noindent 

After Fierz rearrangement of the terms in the 
effective Lagrangian ${\cal L}_{{\rm eff}}$,  
the scales $\Lambda_{n1,n2,c1,c2}$ can be seen as masses of 
effective scalar $SU(2)_{L}$-singlet
spinor bilinears consisting of a 
mirror and an ordinary fermion. These are
reminiscent of ``techniscalars" as to their quantum numbers.  
One may observe that, unlike the 
present situation, the corresponding scalar effective
operators induced by ETC
interactions in ordinary technicolor theories would be
$SU(2)_{L}$-doublets and would not produce shifts to the 
right-handed fermion couplings. The scalar operators appearing here
could in principle correspond to mesons bound with the
QCD force, but their constituents are very heavy and 
are expected in principle to decay weakly before they have time to 
hadronize.

Alternatively, one may think of effective four-fermion operators
of the general form 
${\cal O}(\bar{\psi^{M}_{L}}\psi_{R})(\bar{\psi_{R}}\psi_{L}^{M})$
where the dimensionful form factors ${\cal O}$ are influenced
by non-perturbative effects and are renormalized
differently to lower scales
according to the couplings and masses of the participating
fermions. The operators in question are gauge-invariant and  
on dimensional grounds irrelevant.    
Therefore, the corresponding form factors related to heavier fermions
are in general expected to be more seriously damped at lower scales
than the ones corresponding to lighter fermions, in accordance to 
the decoupling theorem \cite{Tom}.
This fact gives the potential to $S$ to
receive substantial negative contributions, as will become
clear next. 

It is as a matter of fact difficult to predict the values of
the effective couplings of the operators that determine the  
fermion anomalous couplings, since 
they are influenced by non-perturbative dynamics.   
The values of the various terms  
are  here chosen for illustration purposes to be  
$\frac{\lambda_{n1}v^{2}}{\Lambda^{2}_{n1}}=1$,   
$\frac{\lambda_{c1}v^{2}}{\Lambda^{2}_{c1}}=3.32$,   
$\frac{\lambda_{n2}v^{2}}{\Lambda^{2}_{n2}}=0.22$,   
$\frac{\lambda_{c2}v^{2}}{\Lambda^{2}_{c2}}=0.1$. If the strong
mirror group is dynamically broken,
one might wish to 
consider all these operators near criticality
in the Nambu-Jona-Lasinio sense, 
in which case the couplings $\lambda_{n1,c1,n2,c2}$ would be of
order $4 \pi^{2}$. One then finds for example that the smallest scale 
entering this discussion is $\Lambda_{c1} \approx 820$ GeV. 

The terms corresponding to operators involving the standard-model top
quark (see subscripts $n1, c1$ above)
are assumed larger than the ones involving the standard-model bottom 
quark (subscripts $n2, c2$). 
This might be related directly or not to the fact that, as was already seen 
in subsection 3.1,  the masses corresponding to the
$\bar{t_{R}}t^{M}_{L}$ gauge-invariant
 terms, which constitute these four-fermion operators after
 Fierz transformation, are much larger than 
the $\bar{b_{R}}b^{M}_{L}$ terms, because one has to reproduce the
correct top-bottom mass hierarchy (recalling that $m_{t}/m_{b} \approx 35$). 
This is consistent with the  general expectation
in dynamical symmetry breaking schemes of  having 
effects grow larger for heavier fermions, here traceable to the
corresponding larger fermion-mirror fermion mixing. 
This would also explain why four-fermion terms involving first- and
second-generation quarks are neglected in this analysis. 

The possible connection of the
four-fermion operators introduced above with the  mass generation 
process is a subtle issue that could potentially
shed more light on the hierarchy of  
the effective couplings appearing here. With regard to the 
difference between terms with subscripts $n_{i}$ and $c_{i}$, with 
$i=1, 2$, the
ones corresponding to the field $t^{M}_{L}$ are taken here
to be smaller than the ones for $b^{M}_{L}$.
This is motivated by the fact that the mirror-top is much heavier than
the mirror bottom, and it is plausible that the relative
form factors are considerably 
suppressed in comparison with the ones involving the
mirror-bottom. This working assumption will prove
to be crucial for the reported values of the electroweak parameters
in this particular numerical example. 

Using the values above  one finds the anomalous couplings 
 $\delta g^{b}_{R}=-0.03$ and  $\delta g^{t}_{R}=-0.58$.  
The coupling $\delta g^{b}_{R}$ 
is  within  its best-fit experimental 
value $\delta g^{b}_{R}  = 0.036 \pm 0.068$ 
(this is a combined fit including
information on $\delta g_{L}$ and the $S$ and $T$ parameters 
\cite{DoTe}). It is already so tightly constrained 
that, even if it finally turns out to be positive, as suggested by
\cite{Field} and which is 
easily achievable here by an appropriate choice of the
relevant four-fermion couplings, it
will not change our conclusions substantially. 
The coupling $\delta g^{t}_{R}$ is of course not yet tightly constrained, and
it is therefore a good candidate for 
a possible source of the large vertex corrections needed in this model.  

One should expect therefore
that, apart from the
model-independent ``oblique" contributions to the electro-weak 
precision parameters $S$ and $T = \Delta\rho / \alpha$ (where
$\alpha$ is the fine structure constant), denoted by 
$S^{0}$ and $T^{0}$, these parameters   
receive also important vertex corrections 
$S^{t,b}$ and $T^{t,b}$ due to the top and bottom quarks, 
which should be  given in terms of the anomalous couplings 
calculated above. 
The ``oblique" positive corrections to $S$ are given by $S^{0}=0.1N$
for $N$ new $SU(2)_{L}$ doublets, assuming QCD-like strong dynamics. 
On the other hand, the mass
difference  between the up- and down-type mirror fermions 
produces a positive contribution to $T^{0}$. Considerations in the past
literature with mirror
fermions or vector-like models 
which can give very small or negative $S^{0}$ and $T^{0}$ do not 
concern us here because they are, unlike the present case, based on
the decoupling theorem due to the existence of
large gauge-invariant masses \cite{litri}, \cite{many}. 

By summing up these effects therefore, one finds for $S$ and $T$ 
the expressions \cite{DoTe}  
\begin{eqnarray}
S&=&S^{0}+S^{t,b} = 0.1N 
+ \frac{4}{3\pi}(2 \delta g^{t}_{R} - \delta g^{b}_{R})   
\ln{(\Lambda/M_{Z})} \nonumber \\   
 T&=&T^{0}+T^{t,b} = 
\frac{3}{16\pi^{2}\alpha v^{2}}\sum_{i}^{N}(m_{U_{i}^{M}}-m_{D_{i}^{M}})^{2}
+\delta g^{t}_{R}
\frac{3m^{2}_{t}}{\pi^{2}\alpha v^{2}} \ln{(\Lambda/m_{t})}, 
\end{eqnarray}

\noindent where $m_{U_{i}^{M},D_{i}^{M}}$ denote the masses of the
up- and down-type mirror quarks, $N=12$ in the present case, 
and $\Lambda$ is the cut-off, which is expected to be roughly 
equal to the smallest scale appearing in Eq.15, namely
$\Lambda \approx \Lambda_{c1} \approx 820$ GeV. 
The new contributions to $S$ and $T$ reflect modifications
of the $W$ and $Z$ self-energies due to non-standard top-  and
bottom-quark vertices inserted into the relevant loop diagrams.
Note that
these expressions are valid for small anomalous couplings, but they
are used in the following to illustrate the main effect of the new sector
even though the top-quark anomalous coupling is taken to be quite large.
Anyway, it can be assumed that these effects
can be adequately  absorbed 
in  the expressions for the unknown effective four-fermion couplings.
It is also noted that 
contributions to $S^{0}$ and $T^{0}$ from the lepton sector
are calculated assuming Dirac mirror neutrinos. 

Moreover, one has to stress here that no isospin
splitting whatsoever is required {\it a priori}
in the mirror sector in order to get the
top-bottom quark mass hierarchy, since this can be produced by
differences in the gauge-invariant mass submatrices. 
The dynamical generation of this hierarchy  does not lead
to problems with the $T$ parameter, and this can be traced to the  fact that
the fermion 
condensates which break dynamically the electro-weak symmetry are  
distinct from the electro-weak-singlet
condensates responsible for the feeding-down of 
masses to the standard-model fermions. This is contrary to the
usual ETC philosophy. The reason 
this isospin asymmetry is introduced here is only to cancel the large 
negative
contributions to the $T$ parameter coming from the vertex corrections, 
as  will be seen in the following. 

By using the fermion masses and anomalous couplings  
calculated above, one finds that the parameters $S$ and $T$ are given by 
\begin{eqnarray}
S & \approx & 1.2 -0.48 \ln{(\Lambda/M_{Z})}  
\nonumber \\  
T & \approx & 19.4 \times ( 0.88 - 0.58 \ln{(\Lambda/m_{t})}). 
\end{eqnarray} 

\noindent 
The present best-fit  values for the  electroweak parameters  are 
(note that this is again a
combined fit including b-quark anomalous-coupling information
\cite{DoTe})  
\begin{eqnarray}
S & = & -0.40 \pm 0.55 \nonumber \\
T & = & -0.25 \pm 0.46 \;. 
\end{eqnarray} 

\noindent 
It is apparent that, even though these parameters are still consistent
with zero, they can assume non-negligible negative values approaching even -1  
(at $1 \sigma$), so the loop and vertex corrections do not have to
cancel exactly.

One observes moreover  that for cut-off scales $\Lambda$ 
of about 820 GeV,  values for the $S$ and $T$ parameters consistent 
with experiment are feasible, i.e.   $S \approx 0.14$, $T \approx  -0.3$, 
and this is mainly due to the large negative anomalous coupling
$\delta g^{t}_{R}$.
Similar values for the electro-weak precision parameters
could be achieved with smaller anomalous couplings accompanied with
a larger cut-off $\Lambda$. This would lead to lighter 
mirror fermions in order to reproduce the weak scale correctly,
something that would also automatically imply a larger fermion-mirror fermion
mixing,  
but it would have the undesirable effect of increasing the fine tuning
in the model.

It should not be forgotten nevertheless that it is attempted here to
study 
 non-perturbative theories with dynamics
not easily calculable. 
 For instance,
since some of the new fermions introduced have masses close to the scale
$\Lambda$, 
the effective theory is studied very close to the cut-off
where it is expected to lose its accuracy, and the corresponding 
results could be
consequently distorted. 
Moreover, the numbers quoted  are very 
model-dependent and far from having general validity, 
since different assumptions about
the effective couplings and scales involved would lead to different values
for the $S$ and $T$ parameters.  

One should furthermore note that in such a type of calculational schemes
 it seems like an accident  that the $T$
 parameter is so close to zero, since slight deviations in the 
 parameters can shift it to large positive or negative values due
 to the large parenthesis prefactor.  
The large negative contributions to $T$ can be traced to the large 
absolute value of
 $\delta g^{t}_{R}$ which in its turn is needed to cancel the large  
positive contributions to the $S$ parameter. This problem would
be therefore 
less acute if the ``oblique" positive corrections to $S$ were smaller
and a smaller $\delta g^{t}_{R}$ would thus be able to 
accommodate the experimental data. 

One way to achieve this is to note that
the generation group is broken, leading to non-QCD-like strong dynamics.
If this makes the mirror-fermion masses run much slower with 
momentum, it can reduce the positive 
contributions to the $S$ parameter even by a factor of two \cite{ApGe}.
Another way is to have Majorana mirror neutrinos \cite{georgnew},
\cite{georgnew2}. 
In any case, the purpose of the numerical example presented 
is merely to illustrate that theories of this type may 
potentially  produce negative $S$ and $T$ parameters, which in 
general is not easy in dynamical symmetry breaking scenarios. 
Further phenomenological consequences of the model can be found
in \cite{georgnew2}.

\section{Conclusions}

Motivated by several theoretical arguments 
and possibly by some experimental  
indications that there are new physics around
the TeV scale, we extended the gauge sector of the standard model
and its fermionic content in a left-right symmetric context. We argue 
that doubling  the matter degrees of freedom should be considered 
positively if, instead of just burdening the theory with more parameters,
it renders it more symmetric while simultaneously solving several problems
like electroweak radiative corrections,
fine-tuning, fermion mass generation and mixing, possibly 
absence of strong CP violation and eventual unification at very high
energy scales. 

It was shown that the model    
  sets up a precise theoretical framework for the calculation of
fermion mass hierarchies and mixings.
It gives furthermore rise to dynamics which could 
potentially reconcile the $S$- and $T$- parameter
theoretical estimates  with their experimental values without
excessive fine tuning.  Moreover, the doubling  
of the fermionic spectrum it predicts provides decay modes which
should in principle be detectable  
in colliders like $LHC$ and $NLC$ \cite{georgnew2}. This fact, together
with more precise future 
measurements of possible FCNC and anomalous couplings in 
the third fermion generation render the model  experimentally
testable.

Within the present approach, a deeper understanding of the generation of  
the gauge-invariant mass matrices $m$ and the
effective couplings leading to anomalous third-generation
standard-model fermion couplings to the $Z^{0}$ boson is still
needed. This would settle the question on whether the large positive
loop corrections to the $S$ parameter in this model can be adequately
canceled by vertex corrections without unnatural fine-tuning.
Furthermore, a more complete investigation on how  
the mirror generation groups break just after they
become strong is an important open question. 

\noindent {\bf Acknowledgements} \\
I thank the Boston University theory group for their
hospitality during which S. Chivukula pointed out the problem of
Ref.\cite{litri}. I also thank M. Lindner, N. Maekawa and V. Miransky
for helpful discussions at various stages of this study,
and anonymous referees for improving the paper. 
This work was supported by an {\it Alexander von Humboldt Fellowship}.

\end{document}